\documentclass{sigchi-ext}
\usepackage[T1]{fontenc}
\usepackage{textcomp}
\usepackage[scaled=.92]{helvet} 
\usepackage{graphicx} 
\usepackage{balance}  
\usepackage{booktabs} 
\usepackage{ccicons}  
\usepackage{ragged2e} 
\usepackage{subcaption}
\usepackage{makecell}
\usepackage{tabularx}

\usepackage[suppress]{color-edits}
\addauthor{ak}{blue} 
\addauthor{hz}{red} 
\addauthor{kh}{green} 
\addauthor{placeholder}{red}

\def\plaintitle{Towards a Learner-Centered Explainable AI: Lessons from the learning sciences} 
\def\emptyauthor{}

\title{Towards a Learner-Centered Explainable AI}

\numberofauthors{10}

\author{%
  \alignauthor{%
    \textbf{Anna Kawakami\textsuperscript{1}}\\
    \email{akawakam@andrew.cmu.edu} }\alignauthor{%
    \textbf{Nikos Arechiga\textsuperscript{2}}\\
    \email{nikos.arechiga@tri.global} } \vfil \alignauthor{%
    \textbf{Luke Guerdan\textsuperscript{1}}\\
    \email{lguerdan@andrew.cmu.edu} } \alignauthor{%
    \textbf{Matthew Lee\textsuperscript{2}}\\
    \email{matt.lee@tri.global} } \vfil \alignauthor{%
    \textbf{Yang Cheng\textsuperscript{1}}\\
    \email{yanghuic@andrew.cmu.edu} } \alignauthor{%
    \textbf{Scott Carter\textsuperscript{2}}\\
    \email{scott.carter@tri.global} } \vfil \alignauthor{%
    \textbf{Anita Sun\textsuperscript{1}}\\
    \email{ningjins@andrew.cmu.edu} } \alignauthor{%
    \textbf{Haiyi Zhu\textsuperscript{1*}}\\
    \email{haiyiz@andrew.cmu.edu} } \vfil \alignauthor{%
    \textbf{Alison Hu\textsuperscript{1}}\\
    \email{ayhu@andrew.cmu.edu} } \alignauthor{%
    \textbf{Kenneth Holstein\textsuperscript{1*}}\\
    \email{kjholste@cs.cmu.edu}} \vfil \alignauthor{%
    \textbf{Kate Glazko\textsuperscript{2}}\\
    \email{kate.glazko.ctr@tri.global}} \alignauthor {%
    \textbf{1 Carnegie Mellon University}\\
    \textbf{2 Toyota Research Institute}\\
    \textbf{* Denotes equal contribution} }} 
    
\definecolor{linkColor}{RGB}{6,125,233}
\hypersetup{%
  pdftitle={\plaintitle},
  pdfauthor={\emptyauthor},
  bookmarksnumbered,
  pdfstartview={FitH},
  colorlinks,
  citecolor=black,
  filecolor=black,
  linkcolor=black,
  urlcolor=linkColor,
  breaklinks=true,
}

\begin{document}

\CopyrightYear{2022}
\setcopyright{rightsretained}
\conferenceinfo{Human-Centered Explainable AI Workshop, CHI'22}{April  30--May 6, 2022, New Orleans, LA, USA}
\isbn{978-1-4503-6819-3/20/04}
\doi{https://doi.org/10.1145/3334480.XXXXXXX}
\copyrightinfo{\acmcopyright}

\maketitle

\RaggedRight{} 

\begin{abstract}
In this short paper, we 
argue for a refocusing of XAI around \textit{human learning goals}.
Drawing upon approaches and theories from the learning sciences, we propose a framework for the learner-centered design and evaluation of XAI systems. We illustrate our framework through an ongoing case study in the context of AI-augmented social work.
\end{abstract}

\section{Introduction}
Recent years have seen a surge of interest in the question of how AI systems can be made more ``interpretable’’ or ``explainable’’ to humans. 
Yet these terms are used in reference to many disparate goals within the literature \cite{doshi2017towards, lipton2018mythos, miller2019explanation}. For instance, work on interpretability has sometimes focused on enhancing humans’ ability to mentally simulate and predict an AI system’s behavior \cite{lai2019human, lipton2018mythos, poursabzi2021manipulating} or to evaluate counterfactuals \cite{weld2019challenge}. Other work addresses ways to help humans decompose models, to understand their constituent parts (e.g., parameters) and how these parts fit together \cite{lipton2018mythos}. From a human-centered perspective, these design goals can be understood as supporting different \textbf{human capabilities}, each of which may be more or less useful in different real-world contexts. For example, decomposing a model may be useful when debugging an AI system. In a decision-making context, the ability to identify situations that could impact a model's reliability may be more helpful \cite{holstein2020conceptual, mozannar2021teaching}. 

In this paper, we argue that many, if not all, of the design goals in existing XAI research and practice can be productively reinterpreted as human \textbf{learning goals}. Much current XAI research focuses on designing ways to \textit{make models explainable to humans}. By contrast, building upon recent arguments for centering human \textit{understanding} in XAI research \cite{miller2019explanation, vaughan2020human}, we focus on supporting humans in \textit{learning} about particular AI systems and how to work with or around them. Whereas XAI research often aims at communicating information about an AI system instantaneously and with minimal effort on the part of a human recipient, some learning goals may best be met through longer learning engagements or through deliberate practice and feedback \cite{bansal2019beyond, cai2019hello, lai2020chicago,mozannar2021teaching}.

Drawing lessons from the learning sciences—a scientific and design discipline dedicated to the study of human learning and ways to support it in real-world contexts—we explore the implications of adopting a learning-centered lens for the design and evaluation of human-centered XAI. We propose a framework for learner-centered XAI, which integrates and extends existing concepts from the learning sciences. Finally, we present an ongoing case study illustrating how this framework can be applied in practice. 

\section{A framework for learner-centered XAI}
In this section, we propose a framework for the learner-centered design and evaluation of XAI. We describe how three concepts from the learning sciences—backward design \cite{wiggins2005understanding}, participatory design for learning \cite{disalvo2017participatory}, and ``closing the loop’’ \cite{clow2012learning, maclellan2016apprentice}—can help to guide the design of XAI that positions humans as deliberate and continuous learners. The goals of this framework are to (1) offer a systematic process for designing XAI interfaces that target specific \textbf{learning outcomes}, (2) demonstrate how \textbf{context- and stakeholder-specific needs} can be surfaced and addressed during the design process, (3) combine \textbf{participatory and data-driven methods} to support more contextually-relevant XAI designs, and (4) provide a more rigorous approach for \textbf{evaluating the effectiveness} of XAI.

\begin{figure}
    \includegraphics[scale=0.25]{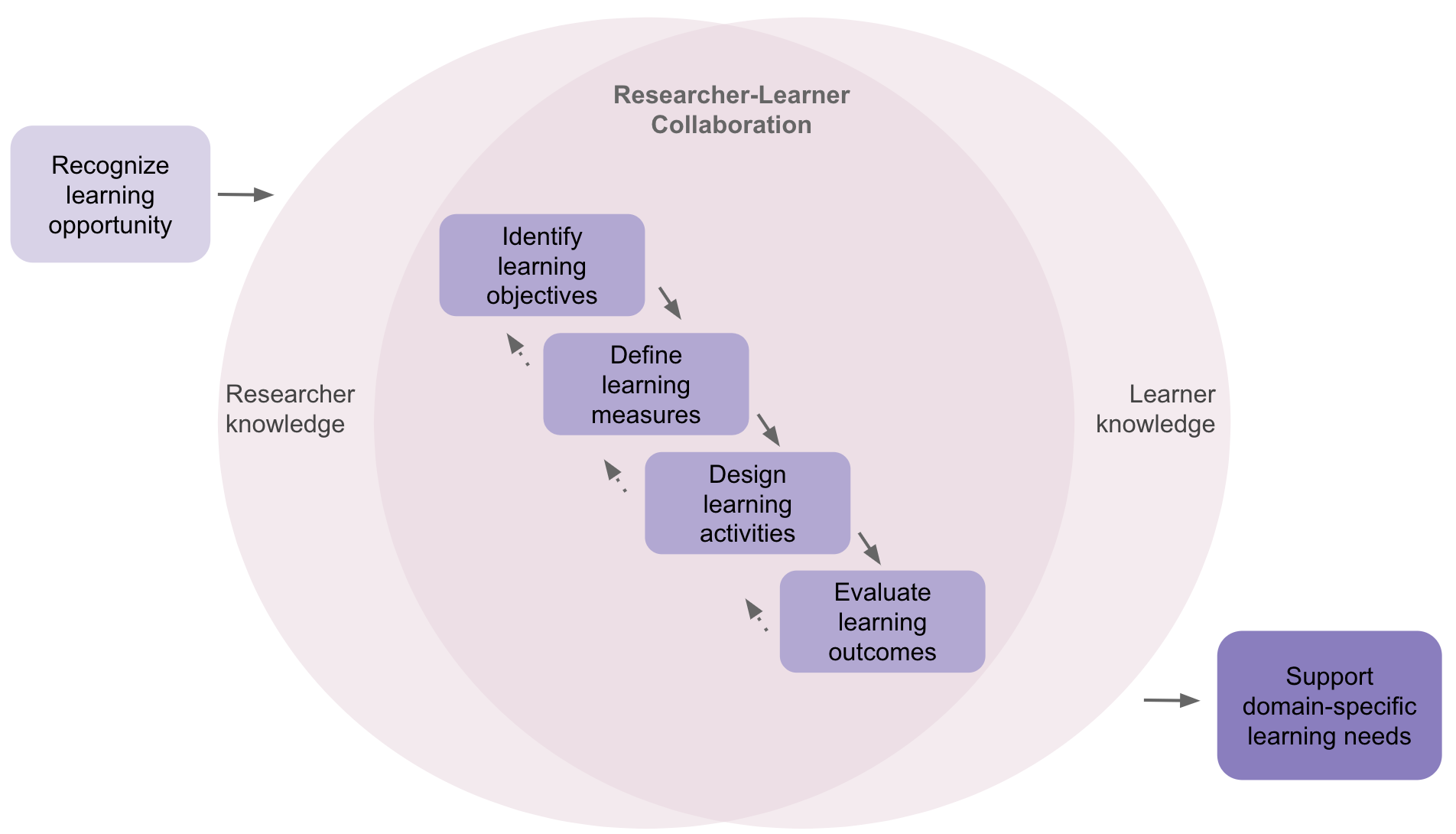}
    \caption{A framework for the design of learner-centered XAI.}
     \label{fig:framework}
\end{figure}

As shown in Figure~\ref{fig:framework}, our framework proposes that researchers should collaborate with relevant stakeholders in real-world human-AI interaction contexts, to iteratively co-design learning objectives, measures, activities, and evaluation approaches. Following a ``backward design’’ approach, as discussed below, this collaboration should begin by specifying \textbf{learning objectives}: a set of specific capabilities that the learners should ideally have following a learning activity. Learners should then be involved in decisions about how to operationalize these learning objectives in the form of concrete \textbf{learning measures} which capture observable human behaviors as proxies for latent constructs such as ``understanding’’ of a targeted concept \cite{jacobs2021measurement}. For instance, researchers might engage learners in specifying \textit{how they would know} whether a given intervention had succeeded in meeting one of their learning objectives: how would they behave differently, or what would they be able to do that they could not do previously? With these objectives and measures in mind, researchers can work with learners to co-design \textbf{learning activities} to try to help learners achieve their specified objectives. The measures specified previously can then be used to \textbf{evaluate learning outcomes}, to guide the iterative, data-driven refinement of learning activities. Below, we introduce three concepts from the learning sciences that inform this framework.

Wiggins and McTighe proposed \textbf{backward design} to address a longstanding challenge in instructional design: teachers and instructional designers often focus more on how to \textit{teach} rather than on how to help students \textit{learn} \cite{aleven2013knowledge,wiggins2005understanding}. Backward design is an approach that `flips’ the design process. Rather than starting with the design of instructional materials, designers are encouraged to first identify desired learning \textit{outcomes}, then to design \textit{assessments} of those outcomes, and lastly to design \textit{instruction} aimed at achieving those outcomes. These challenges in instructional design are echoed in the current XAI landscape: Even as research moves towards more human-centered XAI methods, 
it remains common to first propose an explainability technique, and then evaluate whether and how the technique is useful to users. In the learner-centered XAI framework, we propose a backward design process that \textbf{starts by identifying meaningful learning objectives for a given task context}, then operationalizes what it means to meet those learning objectives. Only after designing and operationalizing learning objectives that reflect stakeholder- and domain-specific needs are XAI designers prepared to design interfaces that meet these learning objectives. 

The framework additionally draws from \textbf{participatory design practices} in the learning sciences. From a learning sciences perspective, participatory design is recast as an opportunity for relevant stakeholders and researchers to collaboratively learn new knowledge that can guide the design process, based on each others’ complementary expertise \cite{disalvo2017participatory}. Stakeholders with relevant lived experience are uniquely positioned to understand their own needs and desires. 
Meanwhile, researchers can bring unique scientific, design, and technical expertise that is critical to designing effective learning interventions. Moreover, as researchers and stakeholders' 
joint understanding of the problem space strengthens, the framework’s emphasis on an \textbf{iterative design process} may encourage them to proactively reflect on their prior design decisions and refine them as needed. Empowering stakeholder participation earlier on in the design process, at the ``defining learning objectives'' stage, not just when evaluating the interfaces, may also open opportunities for different stakeholders in a given context to discuss any misalignments in their envisioned learning needs. 

Finally, Figure~\ref{fig:framework} 
indicates that real-world evaluations of XAI techniques should inform the continuous process of iterative re-design. This aligns with the notion of \textbf{``closing the loop''} in the learning sciences, emphasizing the data-driven refinement of instructional materials based on analysis of data reflecting how people actually learn with them \cite{clow2012learning,maclellan2016apprentice}. This approach offers an opportunity to rigorously evaluate and iterate on co-designed learning objectives, measures, and interfaces, to address design misalignments, or to adapt to changes in stakeholder needs over time.

\section{Case study: Using the framework to design training interfaces for AI-augmented social work}
In this section, we illustrate how the learner-centered XAI framework can be used in practice, through an ongoing case study in the context of AI-augmented social work. 
\subsection{Background}
In an effort to augment social workers’ abilities to efficiently process and prioritize among large volumes of child maltreatment referrals, child welfare agencies have begun to turn to new machine learning-based ADS tools \cite{aclu2021family,Chouldechova2018,saxena2020human,zytek2021sibyl}. The Allegheny Family Screening Tool (AFST) has been in use in Allegheny County, Pennsylvania since 2016, where it assists child maltreatment hotline call screeners and supervisors in prioritizing among referred cases \cite{afstfaq}. While the county has published public-facing reports discussing the ethics and validity of using such a tool \cite{county2017}, recent research raises new concerns around how effectively the tool has been integrated into the organizational and social context in which workers make day-to-day use of the tool.
In particular, in a recent paper, we report findings from a series of interviews and contextual inquiries at this child welfare agency, to understand how workers currently make AI-assisted child maltreatment screening decisions. We found that workers had little to no opportunities to learn about the AI system they were using, nor about how to work with it effectively, limiting their ability to appropriately calibrate their reliance on the tool’s predictions \cite{kawakami2022}. Moreover, we found \khedit{that}\khdelete{deviations between the} workers' decision-making objectives (focusing on short term risks to child safety) \khedit{differed from the}\khdelete{versus the} model's predictive targets (focus\khedit{ed}\khdelete{ing} on \khedit{\textit{much longer-term predictions}}\khdelete{longer-term predictions} of \khedit{indirect \textit{proxies} of} risk)\khedit{.}\khdelete{,} \khedit{While the tool was intentionally designed to complement workers' focus on immediate outcomes, workers were unsure \textit{how} exactly they were meant to integrate the tool's predictions of long-term risk with their own assessments of immediate safety.}\khdelete{raising concerns with current measures of worker-ADS decision-making, for example, which measure alignment of workers' ADS-assisted decisions with whether the ADS predictive target is observed.} 

Overall, \khedit{these prior}\khdelete{the} findings suggest a need to more broadly \khedit{reconsider and} reconceptualize \khedit{what appropriate roles for ADS}\khdelete{the role of ADS} in social work \khedit{might look like}. This reconceptualization necessitates, at \khdelete{the} minimum, finding ways to understand, empower, and integrate worker perspectives in the design of ADS. \akedit{As a first step towards this vision,} \akdelete{Following up on this prior work, }we are currently exploring ways to \akedit{address \khedit{the gap}\khdelete{deviations} between the current design of the AFST and workers' \khedit{beliefs regarding}\khdelete{perceptions of} what effective human-AI decision-making should look like, and how it should be measured. In this ongoing work, we} engage workers in \akedit{the design of training materials, as a means to identify and design worker-centered}\akdelete{designing} learning objectives, measures, and learning activities\akdelete{ that address their needs}.

\khedit{In this project, we do not plan to fully develop or deploy training materials for the AFST specifically. Indeed, based on our findings thus far, we expect that this co-design process will surface needs for fundamentally different kinds of ADS, not just building training interactions around the existing ADS. Rather, we view the AFST context as a rare opportunity to understand workers' learning goals and needs for support in a highly complex, social decision-making context where an ADS has already been in-use for many years (over half a decade). Beyond this context, we plan to explore the generalizability of our findings (e.g., regarding workers' learning goals) to other AI-augmented social decision-making contexts, such as AI-augmented content moderation.}

\subsection{Ongoing case study}
\begin{figure}
    \includegraphics[scale=0.25]{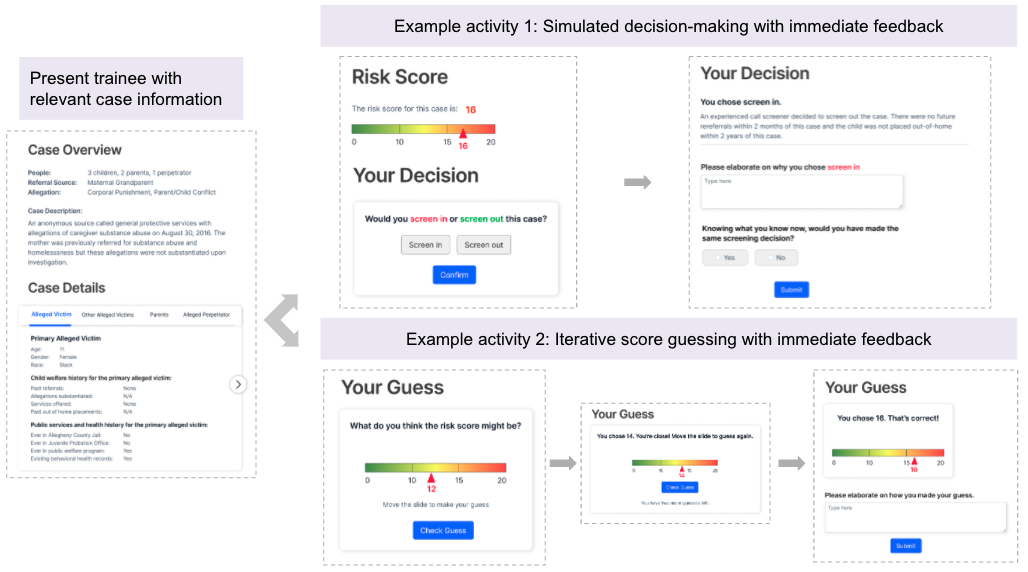}
    \caption{Example interfaces targeting different learning goals.}
     \label{fig:activities}
\end{figure}

Following the first step of the learner-centered XAI framework, we first \textbf{defined a set of fine-grained learning objectives},
such as ``the ability to identify cases where a model may be more or less reliable'',
based on our design research with workers 
(see Appendix\footnote{https://sites.google.com/andrew.cmu.edu/learner-centered-xai/home} for details). 
We plan to further explore, refine, and operationalize these learning objectives in collaboration with various stakeholders (e.g., workers, community members, and agency leadership). 
Figure~\ref{fig:activities} shows two examples of initial training interface sketches that could address specific learning objectives in our taxonomy. In the first example, the learning objective is to improve workers’ ability to appropriately rely on ADS outputs in specific cases. The sketch shows a simulated decision-making activity, which provides low-stakes opportunities for workers to practice integrating their own judgments with AI predictions on real historical data while receiving immediate feedback  \cite{al2010simulation,holstein2020conceptual,koedinger2012knowledge,steadman2006simulation}. The second example sketch focuses on honing workers’ ability to mentally simulate the model’s behavior 
through repeated practice opportunities on a score guessing exercise, with immediate feedback on the closeness of their guesses.  

As next steps, we plan to \textbf{iteratively refine the learning objectives, measures, and activities} through co-design activities with social workers who use this ADS in their daily work, along with other stakeholder groups. Taking a \textbf{participatory design for learning} approach, we view the co-design of learning objectives and measures as an opportunity to surface and address value tensions across different stakeholder groups, regarding what human-AI decision-making in this context should look like in the first place \cite{cai2021onboarding,kawakami2022}. \akedit{For example, while current worker-ADS decision-making performance measures are based on the ADS's predictive target, this assumes the workers should then learn to act like the system would. Our framework aims to involve workers in the design of improved learning measures, to offer alternative measures that counter these assumptions and align more closely with workers' own decision objectives \khedit{or a mixture of workers' decision objectives and the systems' objectives, if that is believed to be desirable}.} \khdelete{After designing and implementing the training activities, we intend to empirically evaluate how well the activities achieve their desired learning objectives, based on the co-designed measures. }

\section{Open Questions}
At the workshop, we hope to further explore several open questions. For example: How might 
learning objectives vary across different human-AI tasks (e.g., prediction, decision-making, or co-creation)? What are other implications of approaching XAI through a learning-centric lens? 



\bibliographystyle{SIGCHI-Reference-Format}
\bibliography{extended-abstract}

\end{document}